\documentclass{gji}
\usepackage{timet}

\usepackage{graphicx}
\usepackage{epstopdf}
\usepackage{amsmath}
\usepackage{amssymb}
\usepackage{color}
\usepackage{caption}
\usepackage{float}
\usepackage[draft]{hyperref}

\title[Atomic clocks as tools to monitor vertical surface motion]
  {Atomic clocks as tools to monitor vertical surface motion}
\author[R. Bondarescu {\it et al.}]
  {Ruxandra Bondarescu$^1$,
  Andreas Sch\"arer$^1$,
  Andrew Lundgren$^2$ \and
  Gy\"{o}rgy Het\'{e}nyi$^{3,4}$,
  Nicolas Houli\'e$^{4,5}$,
  Philippe Jetzer$^1$, 
  Mihai Bondarescu$^{6,7}$
  \\
  $^1$ Department of Physics, University of Z\"{u}rich, Switzerland \\
  $^2$ Albert Einstein Institute, Hannover, Germany \\
  $^3$ Swiss Seismological Service, ETH Z\"{u}rich, Switzerland \\
  $^4$ Department of Earth Sciences, ETH Z\"{u}rich, Switzerland \\
  $^5$ Department of Civil, Environmental and Geomatic Engineering, ETH Z\"{u}rich, Switzerland \\
  $^6$ Department of Physics and Astronomy, University of Mississippi, Oxford, MS, USA \\
  $^7$ Facultatea de fizica, Universitatea de Vest, Timisoara, Romania \\
  }
\date{\today}
\pagerange{\pageref{firstpage}--\pageref{lastpage}}
\volume{???}
\pubyear{2015}

\begin{document}

\label{firstpage}

\maketitle

\begin{summary}

According to general relativity, a clock experiencing a shift in the gravitational potential $\Delta U$ will measure a frequency change given by $\Delta f/f \approx \Delta U/c^2$.  The best clocks are optical clocks. After about 7 hours of integration they reach stabilities of $\Delta f/f \sim 10^{-18}$, and can be used to detect changes in the gravitational potential that correspond to vertical displacements of the centimetre level.
At this level of performance, ground-based atomic clock networks emerge as a tool that is complementary to existing technology for monitoring a wide range of geophysical processes by directly measuring changes in the gravitational potential.
Vertical changes of the clock's position due to magmatic, post-seismic or tidal deformations can result in measurable variations in the clock tick rate.
We illustrate the geopotential change arising due to an inflating magma chamber using the Mogi model, and apply it to the Etna volcano.
Its effect on an observer on the Earth's surface can be divided into two different terms: one purely due to uplift (free-air gradient) and one due to the redistribution of matter.
Thus, with the centimetre-level precision of current clocks it is already possible to monitor volcanoes.
The matter redistribution term is estimated to be $3$ orders of magnitude smaller than the uplift term. Additionally, clocks can be compared over distances of thousands of kilometres over short periods of time, which
improves our ability to monitor periodic effects with long-wavelength like the solid Earth tide.
\end{summary}


\section{Introduction}

\begin{figure}
\centering
\includegraphics[width=8cm]{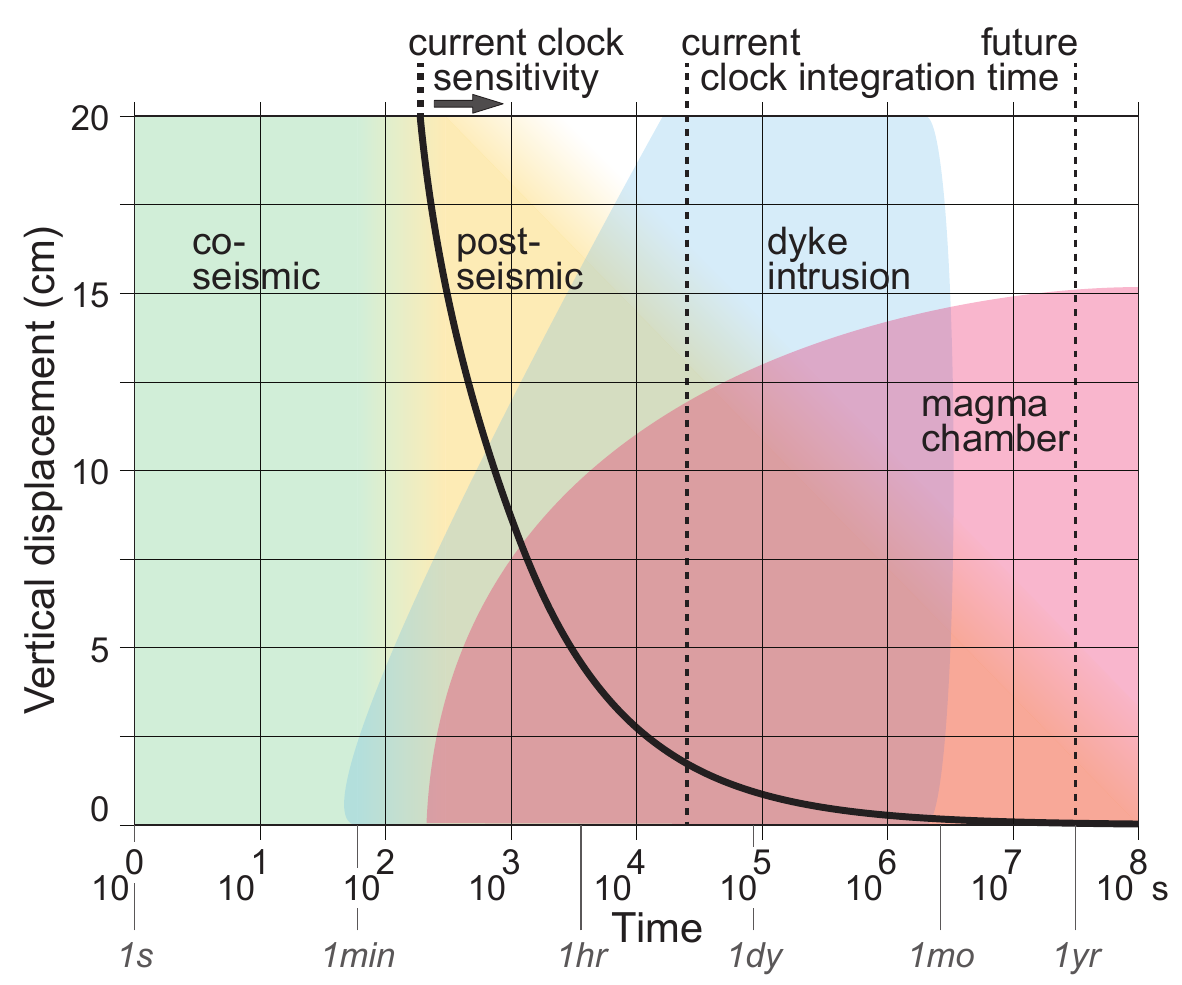}
\caption{Phenomena that could be monitored with optical clock networks. The black
line shows the lower sensitivity of today's best optical clocks; see equation (\ref{eqf2}) and $\Delta z_{\rm today}$ in (\ref{currprec}). The vertical dashed lines show current and planned clock integration times.}
\label{fig:all}
\end{figure}

Vertical deformation transients are key to characterising many geological processes such as magmatic or tectonic deformation (Fig. \ref{fig:all}). Many of these processes have timescales from hours to years which are difficult to measure with existing instruments. Atomic clocks provide a new tool to resolve vertical displacement, with a current precision of about 1 cm in equivalent height after an integration time of 7 hours \cite{Hinkley2013,Bloom2014,Nicholson2015}.

In the past, we argued \cite{Bondarescu2012} that clocks provide the most direct local measurement of the geoid, which is the equipotential surface (constant clock tick rate) that extends the mean sea level to continents. Since a clock network is ground-based, it can provide variable spatial resolution and can be used to calibrate and add detail to satellite maps, which suffer from aliasing (errors due to effects faster than the sampling rate) and from the attenuation of the gravitational field at the location of the satellite.

In this paper, we consider dynamic sources that cause both vertical displacement and underground mass redistribution which produce changes in the local geopotential. Geopotential differences $\Delta U$ are directly measured by the changes in clock tick rate $\Delta f/f \approx \Delta U/c^2$, where $c$ is the speed of light. To be useful, a clock must always be compared to a reference clock, which could be nearby or thousands of kilometres away. Clocks are connected via ultra-precise fiber links that are capable of disseminating their frequency signals over thousands of kilometers with a stability beyond that of the clock \cite{Droste.2013}. As a concrete example we present the case of the inflation (or deflation) of an underground magma chamber, computed analytically \cite{Mogi1958}, and apply it to the Etna volcano. We explore whether the magma filling could be detected using one or two clocks located on the volcanic edifice. 

The primary tools currently used to monitor vertical displacement are InSAR and GPS. Interferometric Synthetic Aperture Radar (InSAR) measures millimetre displacements in the line of sight of radar satellites over wide areas (e.g. B\"urgmann {\it et al.} 2000; Biggs {\it et al.} 2011), but with limited sampling rates (days to weeks). GPS is able to measure vertical displacements of $1$ cm over short timescales ($\sim$ an hour) only when 
the displacement is very localized in the network and/or the frequency of motion is different from the frequency of various artefacts
 that impact GPS accuracy. After surveying areas for more than 10 years, the level of accuracy of GPS techniques is close to the millimetre level \cite{blewitt2002,Houlie.2012} or better, enabling us to better characterize the crustal elastic contrast of plate boundaries \cite{houlie.CA.2009,houlie.CA.2011}.
Since the primary source of noise in GPS measurements is due to signal dispersion through the
atmosphere, both differential GPS and post-processed GPS data
perform better if networks are dense (e.g. Khan {\it et al.} 2010; Houli\'e {\it et al.} 2014) because many artefacts cancel across networks 
over which the ionosphere and troposphere can be assumed to be constant. For timescales of seconds, broadband seismometers can be used \cite{Houlie.2007}, but their bandwidth is unsuited for resolving long-term displacements \cite{boore2003}.

Unlike the GPS or InSAR measurements, local atomic clock measurements are insensitive to atmospheric perturbations, and could resolve ground displacement over shorter integration timescales (hours to months). Further, clocks in conjunction with gravimeters are also able to resolve density changes in the Earth crust that do not, or just partially, lead to uplift or subsidence.  In the case of a spherical magma chamber, the geopotential term resulting from mass redistribution is inversely proportional to the distance to the source $\sim 1/R$, whereas the ground displacement term scales with $1/R^2$. As opposed to this, both terms have the same $1/R^2$ scaling in gravity surveys.
Comparing the measured gravity change to the uplift, $\Delta g/\Delta h$, can reduce this degeneracy to model processes a volcano is undergoing before eruption \cite{rymer}and also more general processes (see Fig. 1 and De Linage {\it et al.} 2007).

Even in areas without active seismic or volcanic processes, the solid Earth tide has a vertical amplitude that can be as high as 30 cm \cite{agnew} with a semi-diurnal period, whose amplitude can be monitored by an atomic clock that is compared with a distant reference clock. We find that geopotential and gravity measurements are sensitive to two different combinations of tidal Love numbers, and could be used to calibrate existing measurements of the solid Earth tide.


\section{Overview of Atomic Clocks}

According to Einstein's theory of general relativity, time slows down in the vicinity of massive objects.
On a neutron star, clocks tick at about half their rates on Earth.
An observer outside the horizon of a black hole even sees time stopping all together at the horizon.
Similarly, clocks that are closer to Earth tick slightly slower than clocks that are further away.

Atomic clocks employ atomic transitions that are classified, depending on the transition frequency, as either microwave or optical clocks. Since the clock frequency depends only on a known atomic transition and constants of nature, clocks respond identically 
to changes in the gravitational potential and do not require calibration. This is in contrast to relative gravimeters, which suffer from instrument drift and have to be calibrated via comparisons with other measurements at the same location.
The current definition of the second is based on a microwave atomic clock. However, optical clocks have the potential for higher stability because they utilize atomic transitions with resonance line-widths typically $10^{5}$ narrower than microwave transitions.  
Since the development of the femtosecond laser frequency comb, optical clocks have been improving extremely rapidly \cite{Poli2013}.
Today's best clocks are optical. They are laboratory devices with frequency uncertainty
\begin{align}
 \Delta f/f \sim 3 \times 10^{-16}/\sqrt{\tau/\rm{sec}},
 \label{eqf2}
 \end{align}
where $\tau$ is the integration period  \cite{Hinkley2013,Bloom2014,Nicholson2015}, and are likely to continue to improve dramatically within the next decade \cite{Poli2013}.
A transportable optical clock that monitors and compensates for environmental effects
(temperature, pressure, electric and magnetic fields) and fits within two cubic meters has recently been built \cite{Poli.2014transportable}.

In the future, it is expected that clocks will become sensitive to surface displacements at the sub-millimetre level.
Possible techniques include nuclear optical transitions \cite{Campbell2012}, 
optical transitions in Erbium \cite{Kozlov2013}, and transitions in highly charged ions \cite{Derevianko2012}.
A stability of 
\begin{align}
\Delta f/f \sim \sigma_{\rm tomorrow} = 10^{-17}/\sqrt{\tau{\rm/sec}}
\label{future}
\end{align}
 should be possible \cite{Hinkley2013}, which would achieve $\Delta f/f \sim 10^{-20}$ within one month.

In order to take full advantage
of the improved stability of optical clocks, distant clocks have to be compared reliably to the $10^{-18}$ level. This entails a global understanding of vertical displacements, the solid Earth tide, and, overall, of the geoid
to the $1$-cm level. Any effects that cause perturbations to this level would have to be reliably modelled and understood.
Clock comparisons via satellites are currently limited by the precision of the communication link that passes through a potentially turbulent atmosphere.
The most precise comparisons of distant clocks currently use underground optical fiber links.
Optical frequency transfer with stability better than the clock has been demonstrated over a two-way distance of 1840 km \cite{Droste.2013}, with a fiber-link from Braunschweig, Germany to Paris, France.
A fiber-link network capable of disseminating ultra-stable frequency signals is being planned throughout Europe (e.g. NEAT-FT collaboration, REFIMEVE+). 

\section{Methods}

While clocks are sensitive to changes in the gravitational potential, relative gravimeters see changes in the vertical component of the gravitational acceleration, which is a vector ($\vec{g} = - \vec{\nabla} U$) whose amplitude can generally be measured much better than its direction.
They both provide local measurements of the change in gravity or potential relative to a reference point where the gravity and potential are known accurately.
Within a large fiber links network, the reference clock could be very far away.

For a displacement $z$, the geopotential and gravity changes are
\begin{align}
\Delta U  \approx - \frac{G M_\oplus}{R_\oplus^2} ~ z \label{pot}, \; \Delta g \approx \frac{2 G M_\oplus}{R_\oplus^3} ~ z.
\end{align}
where $G$ is the gravitational constant, and $M_\oplus$ and $R_\oplus$ the mass and radius of the Earth.
The well-known free-air correction $\Delta g$ for gravity is the gradient of the potential change $\Delta U$.

A vertical displacement of  $z = 1$ cm causes changes in the geopotential and in gravity of
\begin{align}
\Delta U & \approx 0.1 \left(\frac{z}{1 \; \rm cm}\right) \frac{\rm m^2}{\rm sec^2}, \\ 
\Delta g  & \approx 3 \times 10^{-8} \left(\frac{z}{1 \; \rm cm}\right) \frac{\rm m}{\rm sec^2} \sim 3 \left(\frac{z}{1 \; \rm cm}\right)\, \mu {\rm gal}.
\end{align}
Thus, the frequency of a clock changes by
\begin{align}
\label{eqf1}
\frac{\Delta f}{f} \approx \frac{\Delta U}{c^2} \sim 10^{-18} \left(\frac{z}{1 \; \rm cm}\right).
\end{align}

The solid line in Fig. \ref{fig:all} is the vertical displacement $\Delta z_{\rm today}$ as a function of 
clock integration time $\tau$, which is obtained by equating (\ref{eqf2}) and (\ref{eqf1}), while $\Delta z_{\rm tomorrow}$ results
from (\ref{future}) and  (\ref{eqf1}).
\begin{align}
\label{currprec}
 \Delta z_{\rm today} \approx 300 \left(\frac{\tau}{\rm sec}\right)^{-1/2} \rm{cm}, \; \Delta z_{\rm tomorrow} \approx 10 \left(\frac{\tau}{\rm sec}\right)^{-1/2} \rm{cm}.
\end{align}


\section{Applications}

We first discuss the geopotential and gravity changes caused by an inflating magma chamber, using the Mogi model.
This is followed by a discussion of the measurability of the solid Earth tides.

\subsection{Inflating magma chamber - the Mogi model}
\label{sec:Inflating magma chamber}

An inflating or deflating magma chamber can be described by the so-called ``Mogi model'' \cite{Mogi1958}: an isolated point pressure source in an elastic half-space that undergoes a pressure change. The Mogi model is broadly used in the literature (e.g. Houli\'e {\it et al.} \shortcite{Houlie.etna} and Biggs {\it et al.} \shortcite{Biggs.2011}).
Recently, its predictions were compared to sophisticated simulations, which showed that in many situations the discrepancy is quite small \cite{Pascal}.

Clocks and gravimeters lying above the magma chamber will be affected by
(1) a ground displacement, (2) the change of mass density within the chamber, (3) the uplifted rock, and (4) the change in the density of the surrounding material.
We refer to effects (2), (3) and (4) as `mass redistribution' ($\Delta U_2$, $\Delta U_3$ and $\Delta U_4$ in 
Appendix A). The change of the gravitational potential due to ground displacement (1) arises since the observer is shifted to a position farther away from the centre of the Earth. A Mogi source centred at $(0,d)$ that undergoes a volume change $\Delta V$ deforms the half-space, lifting an observer sitting at $(r,0)$ by
\begin{align}
|w| & = (1-\nu) \frac{\Delta V}{\pi} \frac{d}{(r^2 +d^2)^{3/2}},
\end{align}
where $\nu$ is Poisson's ratio. For crustal rock the typical value is $\nu \approx 0.25$.
Using Eq.\ \eqref{pot} with $z = w$, this uplift changes the potential by
\begin{align}
\begin{split}
\Delta U_1 & \approx - \frac{G M_\oplus}{R_\oplus^2} w
= \frac{G M_\oplus}{R_\oplus^2} (1-\nu) \frac{\Delta V}{\pi} \frac{d}{(r^2+d^2)^{3/2}},
\end{split}
\end{align}
where higher order terms have been neglected.
When we add the contribution of the mass redistribution terms
\begin{align}
\Delta U_\text{m} = G \rho_m \Delta V \frac{1}{\sqrt{r^2+d^2}}.
\end{align}
The total geopotential change is
\begin{align}
\begin{split}
\Delta U &= \Delta U_1 + \Delta U_\text{m}
\\
&= \frac{G M_\oplus}{R_\oplus^2} (1-\nu) \frac{\Delta V}{\pi} \frac{d}{(r^2+d^2)^{3/2}} + G \rho_m \Delta V \frac{1}{\sqrt{r^2+d^2}},
\end{split}
\end{align}
where $\rho_m$ is the magma density.
Typically, the change due to mass redistribuiton $\Delta U_{\rm m}$ is several orders of magnitude smaller than that due to uplift $\Delta U_1$.
Measuring the mass redistribution term would require that we subtract the uplift term which may be obtained individually using other techniques like GPS or InSAR.

Similarly, the change in the gravitational acceleration is given by
\begin{align}
\begin{split}
\Delta g  &= \Delta g_1 + \Delta g_m
\\
&= \left[- \frac{2 G M_\oplus}{R_\oplus^3} \frac{1-\nu}{\pi} + G \rho_m \right]
\Delta V \frac{d}{(r^2+d^2)^{3/2}}.
\end{split}
\end{align}

\begin{figure}
  \centering
  \includegraphics[width=8cm]{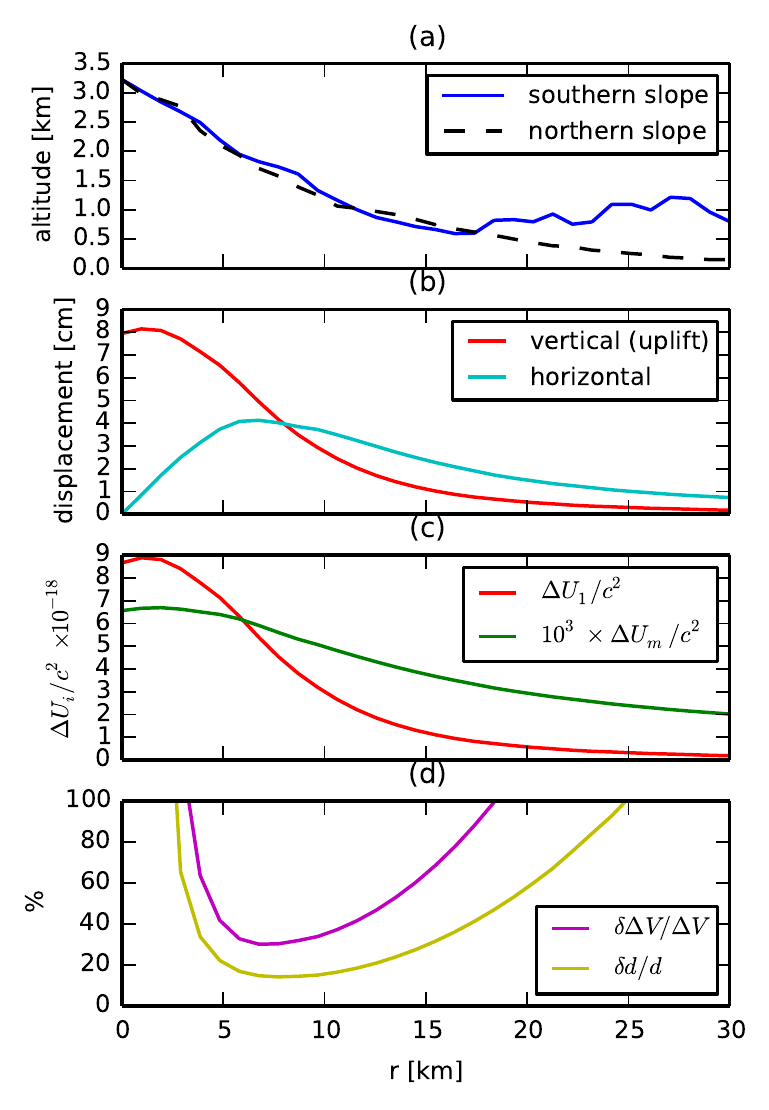}
  \caption{Estimate of vertical ground deformation on the Etna volcano over the course of one year with the Mogi model.
  In (a), the altitude profile of Etna is shown. The solid curve shows the southern slope of the mountain, the dashed line shows the mirror image of the northern slope. For the following plots only the former is considered since the latter gives very similar results.
  Assuming a Mogi source located $9.5$ km below the summit with a volume change rate of $30\times 10^6 \text{m}^3/\text{yr}$, the vertical and horizontal motion over one year is plotted in (b). The change in the gravitational potential due to this uplift together with the change due to mass redistribution is shown in (c). Notice that the latter is about $3$ orders of magnitude smaller.
  If one clock is positioned at the summit, (d) shows the fractional errors on the measurements of the source depth $\delta d/d$ and the volume change $\delta\Delta V/V$ as a function of the horizontal distance to a second clock.
  }
\label{fig:etna}
\end{figure}

Notice that for both the acceleration and the potential the different terms of (2), (3) and (4) mostly cancel each other and only a term $\sim \rho_m$ survives (see Appendix A for the complete calculation).
However, this cancellation is a consequence of the elastic half-space assumption of the Mogi model and will be less exact in more realistic scenarios.
On the one hand, for the acceleration all terms have the same spatial dependence.
On the other hand, for the potential they scale differently: while the uplift term scales as $d/R^3$ like all acceleration terms, the mass redistribution term scales as $1/R$, where $R = \sqrt{r^2 +d^2}$ is the distance to the source.

Assuming a clock with stability $\Delta f/f = \sigma_0/\sqrt{\tau/{\rm sec}}$ and a source with a constant volume change rate $\Delta V/\Delta \tau$, the minimum required integration time to measure the uplift at a location $(r,d)$
\begin{align}
\label{equ: integration time uplift}
\tau_1 = \left[ \frac{ \sigma_0 \pi c^2}{g (1- \nu)} \frac{R^3}{d} \left( \frac{\Delta V}{\Delta \tau} \right)^{-1} {\rm sec}^{-1} \right]^{2/3} \rm{sec},
\end{align}
with $g = G M_\oplus/R_\oplus^2$, whereas for the mass redistribution
\begin{align}
\label{equ: integration time mass redistribution}
\tau_m = \left[ \frac{\sigma_0 R c^2}{G \rho_m} \left( \frac{\Delta V}{\Delta \tau} \right)^{-1} {\rm sec}^{-1}\right]^{2/3} \rm {sec}.
\end{align}

We focus on the specific example of the Etna volcano.
Houli\'e {\it et al.} \shortcite{Houlie.etna} used GPS data to investigate its underground magma system and found that a Mogi source located about $9.5$ km below the summit with a volume change rate of $30 \times 10^6\, \text{m}^3/\text{yr}$ would yield the observed uplift. 
In Fig. \ref{fig:etna}, we plot the ground motion and the resulting potential change as well as the potential change due to mass redistribution.
We find that $\Delta U_m \sim 10^{-3} \Delta U_1$, which roughly corresponds to $\Delta f/f \sim \Delta U/c^2 \sim 10^{-20}$.
If we assume constant volume change rate, an optical clock on the summit would see the uplift if integrated for about ten days (Eq. \eqref{equ: integration time uplift}), 
while the mass redistribution signal is out of reach. With a clock stability $\sigma_{\rm tomorrow}$, such an uplift would 
be observable within a day, and the mass redistribution in five months. 

We also give an example of how to best choose the clock locations. We assume that the horizontal position of the magma chamber is already known from previous surveys, and that the two measurements we wish to make are the depth of the magma chamber $d$ and the volume change $\Delta V$ (assuming also that $\nu$ is known). With one clock directly above the magma chamber, we find the location of a second clock which minimizes the measurement errors on $d$ and $\Delta V$; of course there must also be a distant reference clock.
For a flat half-space we find the optimal location to be at a horizontal
distance $r \approx 0.78 d$ (note that the minima is broad; see Appendix \ref{optimalClock}).
Assuming Etna to be a half-space with chamber depth $9.5$ km, the fractional errors there are $\delta d/d \approx 18 \%$ and $\delta \Delta V/\Delta V \approx 38 \%$.
Including the height profile of the mountain, Fig. \ref{fig:etna}(d) shows $\delta d/d$ and $\delta \Delta V/\Delta V$ as a function of the distance to the summit, assuming a clock sensitive to $\Delta U/c^2 = 10^{-18}$. We find that the optimal location of a second clock is $r \approx 0.78 \, d$, independent of the clock's performance. For the performance considered here, this corresponds to $\delta d/d \approx 14 \%$ and $\delta \Delta V/\Delta V \approx 30 \%$.


\subsection{Solid Earth Tides}

Solid Earth tides are the deformation of the Earth by the gravitational fields of external bodies, chiefly the Moon and the Sun.
In general relativity the Earth is freely falling, so at the centre of the Earth the external gravitational force is canceled by the Earth's acceleration toward the external body. Because the external field is not uniform, other points experience a position-dependent tidal force \cite{agnew}.

The tide has three effects of concern on the gravitational potential. First is the external potential itself, which can be calculated directly from the known mass and position of the external bodies. Second is the change in the Earth's gravitational potential produced by the deformation of the Earth. Third is the vertical displacement of the surface, which produces a free-air correction to our measurements. The last two effects are proportional to the first, parameterized by the Love numbers $k_n$ and $h_n$, respectively \cite{agnew}.

As shown in Appendix \ref{subsec:Solid Earth tides}, clocks (sensitive to geopotential changes) and gravimeters (sensitive to the downward component of $g$) each measure a different linear combination of the Love numbers. This is because the three effects scale differently with the distance from the centre of the Earth. It is therefore desirable to combine gravimeter and clock measurements to infer the Love numbers with great accuracy.
Recall that clock measurements must be made between a pair of clocks. Since tidal effects are global, both clocks are sensitive to the tidal deformation; one clock cannot simply be treated as a reference.
To measure the tidal amplitude, it is necessary to compare instruments over distances of the order of half the tidal wavelength on timescales shorter than the period of the tide. The tidal wavelength is half the circumference of the Earth for the dominant tidal mode.
Both differential GPS and InSAR have short baseline, so tidal effects are small and are mostly subtracted by modelling.
A network of clocks where each clock pair is separated by hundreds of kilometers could more holistically monitor solid Earth tides. Current clocks provide measurements 
of the vertical uplift to within a percent of the maximal tidal amplitude on an hourly basis. Such measurements could
be used to monitor stress changes within the crust, and to investigate whether these correlate with triggered seismicity.

Such a network of clocks on the continent scale could accurately measure the tidal Love numbers.
The external tidal potential can be decomposed into a sum of Legendre polynomials
\begin{align}
U_{\rm tid}^n(R, \alpha) = - \frac{G M_{\rm ext}}{R} \left(\frac{R_\oplus}{R}\right)^n P_n(\cos \alpha),
\end{align}
where $R$ is the distance between the external body of mass $M_{\rm ext}$ and the Earth's centre of mass, and $\alpha$ is the angle from the observation point to the line between the centre of the Earth and the external body. Since $R_\oplus/R \approx 1/60$ for the Moon, and $R_\oplus/R \approx 1/23000$ for the Sun, it is typically sufficient
to just consider the first few terms of the expansion. By linearity, we treat each $n$ separately.

The potential change measured by a clock, including all three effects above, is
\begin{align}
\Delta U_n &= (1+k_n-h_n) U^n_\text{tid}(R,\alpha) ~.
\end{align}
The change in the vertical gravitational acceleration measured by a gravimeter is
\begin{align}
\Delta g_n = - n \left(1 - \frac{n+1}{n} k_n + \frac{2}{n} h_n \right) \frac{U_\text{tid}^n(R,\alpha)}{R_\oplus} ~.
\end{align}
Combining the two measurements with the known $U^n_\text{tid}(R_\oplus,\alpha)$ we can determine the Love numbers
\begin{align}
k_n &= \frac{n+2}{n-1} + \frac{R_\oplus \Delta g_n - 2 \Delta U_n}{(n-1) U^n_\text{tid}(R_\oplus,\alpha)},
\\
h_n &= \frac{2 n+1}{n-1} + \frac{R_\oplus \Delta g_n - (n+1) \Delta U_n}{(n-1) U^n_\text{tid}(R_\oplus,\alpha)}.
\end{align}
Both Love numbers are believed to be modelled to within a fraction of a percent \cite{Yuan.2012}. These models will be throughly tested
once a global optical clock network becomes available.

\section{Conclusions}
We have demonstrated the promise of very precise atomic clocks for geophysical measurements with two illustrative examples.
In the inflation or deflation of a spherical magma chamber (the Mogi model) clocks are primarily sensitive to the local vertical displacement resulting at the Earth's surface. Such monitoring of local deformations can be done using a reference clock anywhere outside the zone where the displacement is significant, typically tens of kilometres. However, when monitoring solid Earth tides, the best accuracy will be obtained with a clock network spanning the globe.

Beyond the examples given, it should be possible to use clocks in conjunction with gravimeters to monitor dyke intrusion, post-seismic deformation, aquifer variations, and other effects causing vertical displacement or subsurface mass change.
In contrast to GPS and InSAR measurements, ground clock measurements are insensitive to the turbulence in the atmosphere.

In the future, optical clocks are expected to become part of the global ground-clock network that is used for telecommunications increasing the precision with 
which we can monitor time. Ground clocks can thus be combined with existing instrumentations (GPS, InSAR, gravimeters) to track underground mass redistribution
through its effects on the geopotential. Portable optical clocks have already been developed \cite{Poli.2014transportable}, and could monitor changes in the 
geopotential of the Earth across fault lines, in areas with active volcanoes, and for surveying.


\begin{acknowledgments}
We acknowledge support from the Swiss National Science Foundation, and thank Domenico Giardini for helpful discussions.
\end{acknowledgments}


\pagebreak

\appendix

\section{The Mogi model in detail}

\label{The Mogi model in detail}

A Mogi source is an isolated pressure point source embedded in an elastic half-space.
A pressure change $\Delta P$ deforms the surrounding half-space and uplifts an observer standing on the surface. This ground displacement, discussed in Sec. \ref{sec:Inflating magma chamber}, affects the ticking rate of a clock since its position in the Earth's gravitational field is slightly shifted.
Besides that, there are three additional effects affecting both the gravitational potential and acceleration, which are due to the redistribution of matter.
These are discussed in Sec. \ref{sub:Direct signal}, \ref{sub:The potential of the uplifted rock} and \ref{subsec:Gravitational potential due to density change}.

\begin{figure}
  \centering 
 \includegraphics[width=8cm]{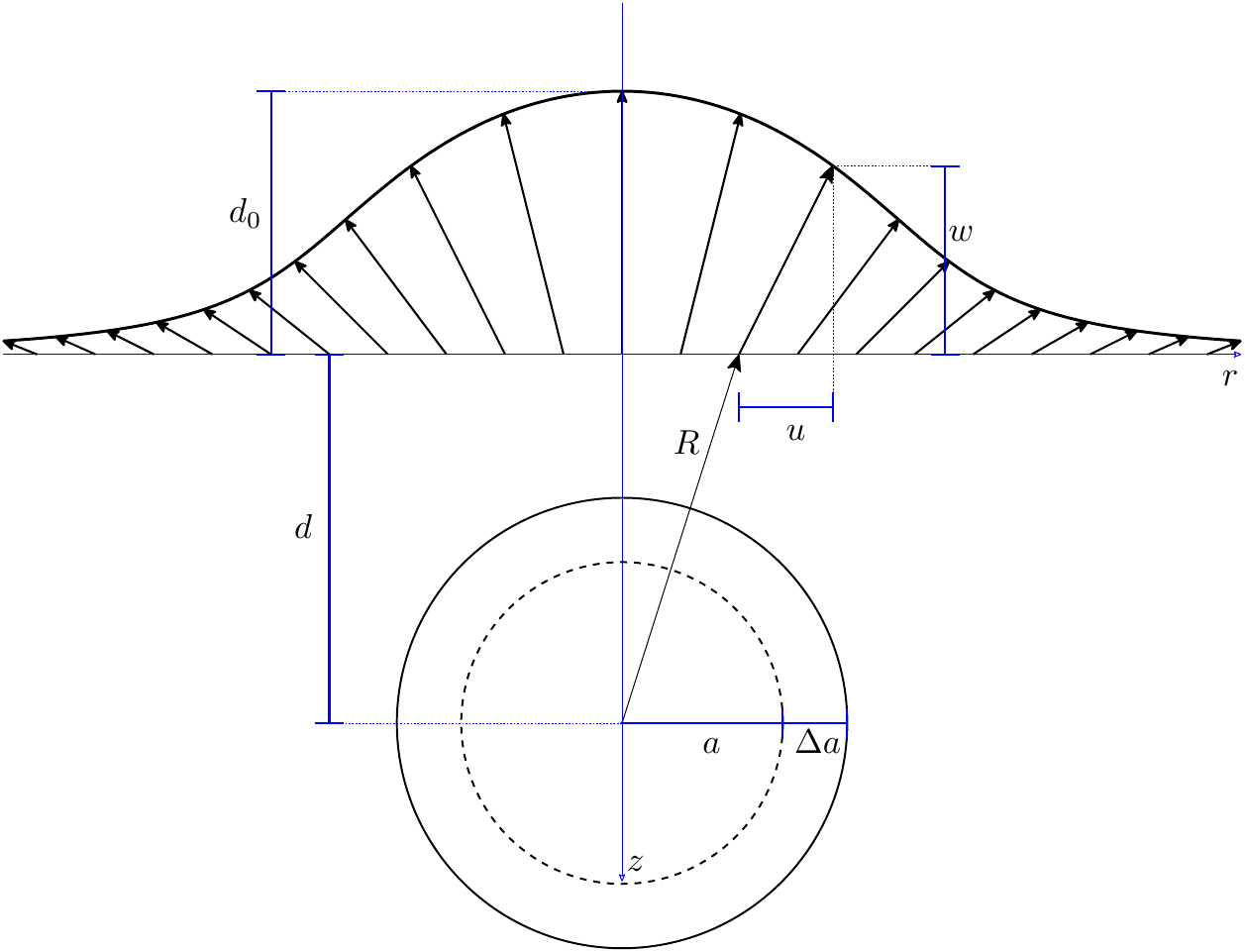}
  \caption{Deformation due to inflating cavity.}
\label{fig:deformation}
\end{figure}

We choose coordinates such that the elastic half-space extends from minus to plus infinity in both the $x$ and $y$ directions, and from $0$ to $+\infty$ in the $z$ direction.
The source is located at $(0,0,d)$ and an observer positioned on the surface has coordinates $(x,y,0)$.
By symmetry, we typically use coordinates $(r,z)$, where $r \equiv \sqrt{x^2+y^2}$ is the distance from the observer to the place on the surface directly above the cavity.
A scheme is given in Fig. \ref{fig:deformation}.

\begin{figure}
  \centering 
 \includegraphics[width=8cm]{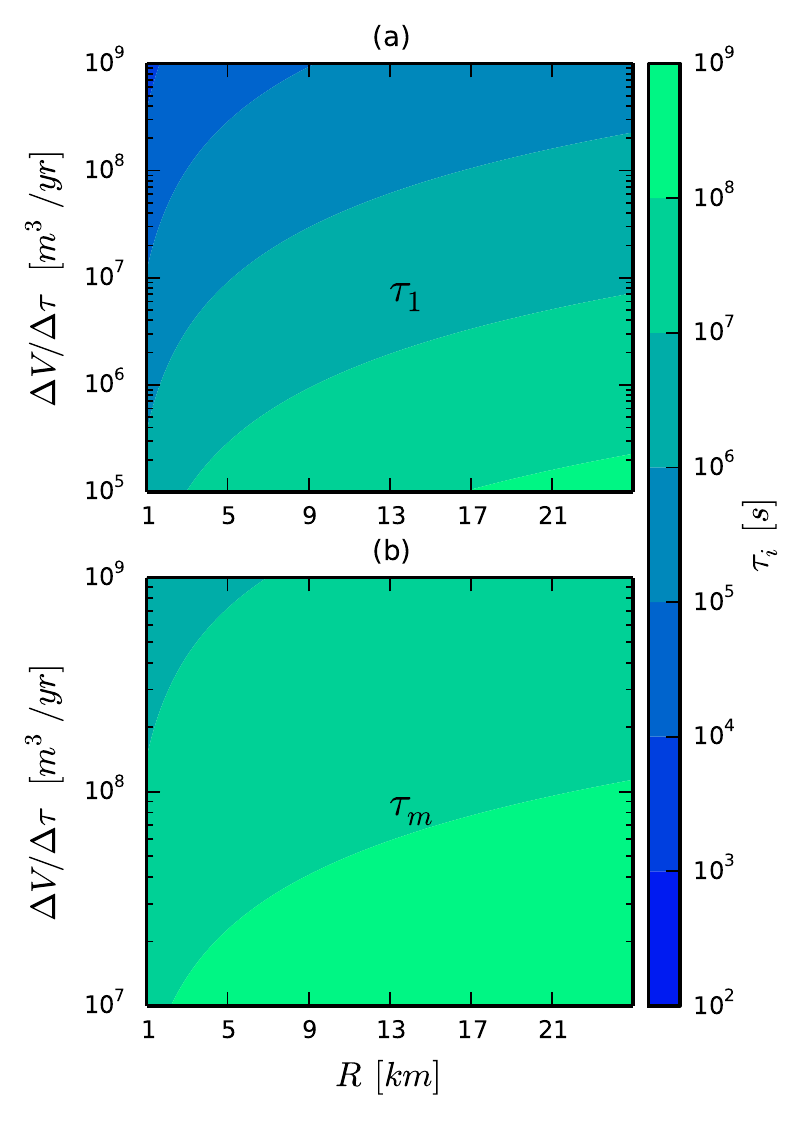}
  \caption{Integration times for current clocks.
  Considering a clock with $\sigma_0 = 3\times 10^{-16}$, the required integration time to resolve uplift (Eq. \eqref{equ: integration time uplift}) is shown in (a) as a function of distance $R$ and (constant) volume change rate $\Delta V/\Delta \tau$.
  Here, it is assumed that the clock is placed above the magma chamber ($r=0$).
  Analogous, the integration time for seeing the redistribution of mass (Eq. \eqref{equ: integration time mass redistribution}) is shown in (b).
  In contrast to the uplift where the result depends on both $r$ and $d$, the integration time for the mass redistribution term depends on the total distance $R$ only.
  }
\label{fig:timescales_current}
\end{figure}

\begin{figure}
  \centering 
 \includegraphics[width=8cm]{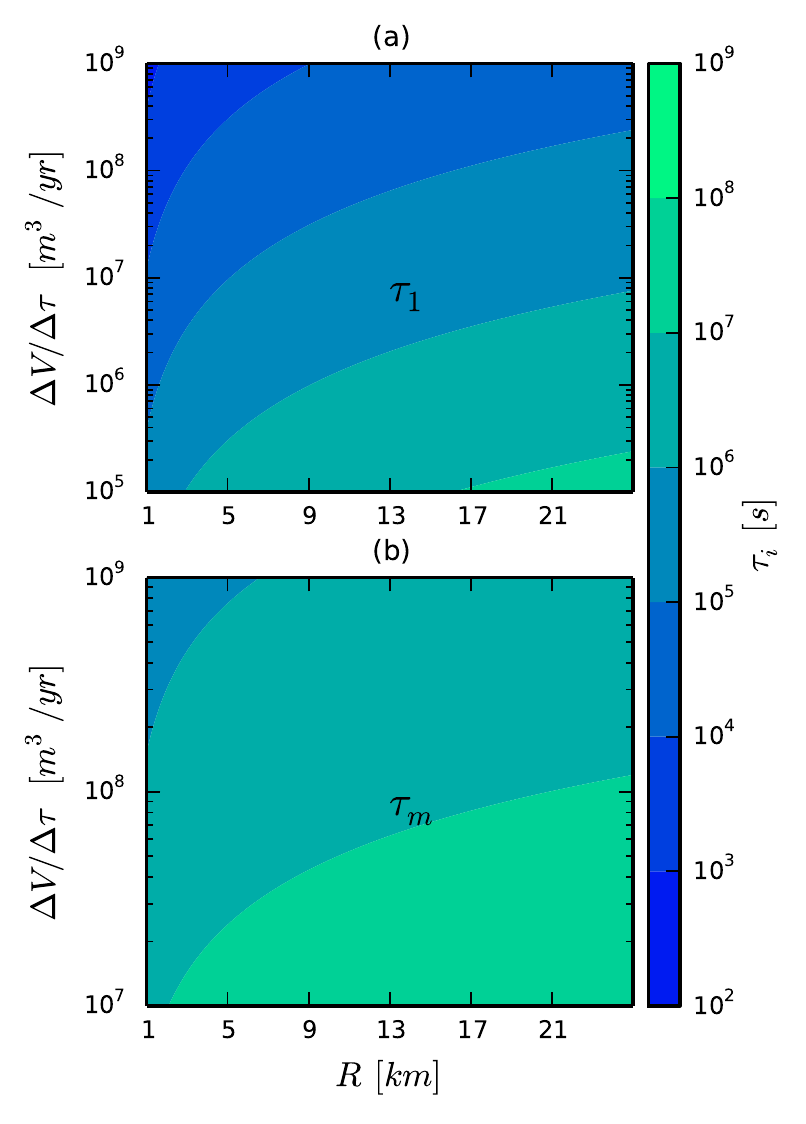}
  \caption{Integration times for future clocks.
  Analogous to Fig. \ref{fig:timescales_current}, the integration times to resolve uplift and mass redistribution are shown for future clocks with $\sigma_0 = 1\times 10^{-17}$.
  }
\label{fig:timescales_future}
\end{figure}

Even though a point source is considered, the pressure change can be interpreted as the volume change $\Delta V$ of a finite size cavity. This approximation is valid as long as the radius of the cavity is much smaller than the depth $d$.
Such a volume change can, for example, be caused by the inflation or deflation of a magma chamber.

The radius of the cavity changes by \cite{Hagiwara1977}
\begin{align}
\Delta a = \frac{a \Delta P}{4 \mu},
\end{align}
where $\mu$ is the shear modulus, having the same units as pressure (note that some authors use $G$ for the shear modulus).
Typical values for rock are $\mu = 10\times 10^{9}$ Pa to $30\times 10^{9}$ Pa.
Thus, the volume changes by
\begin{align}
\Delta V = \frac{4\pi}{3}(a + \Delta a)^3 - \frac{4\pi}{3} a^3 = 4\pi a^2 \Delta a + \mathcal{O}(\Delta a^2).
\end{align}
Neglecting terms of order $\mathcal{O}(\Delta a^2)$, it can be written as
\begin{align}
\label{equ:volume change from pressure change}
\Delta V = \frac{ \pi a^3 \Delta P}{\mu} = \frac{3 V \Delta P}{4 \mu}.
\end{align}
The change in pressure within the cavity induces a volume change that affects the surrounding elastic medium. An observer on the surface at $(r,z=0)$ is displaced by
\begin{align}
\label{equ:surface displacements}
\begin{pmatrix} u \\ w \end{pmatrix}
&= a^3 \Delta P \frac{1-\nu}{\mu} \begin{pmatrix} r \\ -d \end{pmatrix}/R^3 
= (1-\nu) \frac{\Delta V}{\pi} \begin{pmatrix} r \\ -d \end{pmatrix}/R^3,
\end{align}
with $\nu$ being the Poisson ratio and $R \equiv \sqrt{r^2 + d^2}$ being the distance between the cavity and the observer.
Thus, a point on the surface that was originally at $(r,0)$ gets shifted to $(r+u,w)$.
Therefore, an observer on the Earth surface is vertically uplifted by
\begin{align}
\begin{split}
\label{equ:vertical uplift}
|w(r)| &= (1-\nu) \frac{\Delta V}{\pi} \frac{d}{(r^2+d^2)^{3/2}}.
\end{split}
\end{align}

In addition, the density of the elastic body changes at each point by \cite{Hagiwara1977}
\begin{align}
\label{equ: density change}
\begin{split}
\Delta\rho (r,z) &= \frac{\rho a^3 \Delta P}{\lambda+\mu} \frac{r^2 - 2(z+d)^2}{[r^2 + (z+d)^2]^{5/2}}
\\
&= (1-2\nu) \rho \frac{\Delta V}{\pi} \frac{r^2 - 2(z+d)^2}{[r^2 + (z+d)^2]^{5/2}},
\end{split}
\end{align}
where $\lambda$ is Lam\'e's constant
\begin{align}
\lambda \equiv \frac{2\mu \nu}{1-2\nu} \stackrel{\nu = \frac{1}{4}}{=} \mu.
\end{align}

Below, we calculate the changes in the clock and gravimeter measurements due to ground displacement and the three effects coming from mass redistribution. Summing the results of Appendix \ref{sub:Direct signal}, \ref{sub:The potential of the uplifted rock} and \ref{subsec:Gravitational potential due to density change} we find
\begin{align}
\label{equ:potential change mass redistribution}
\begin{split}
\Delta U_\text{m} &\equiv \Delta U_2 + \Delta U_3 + \Delta U_4
= - G \rho_m \Delta V \frac{1}{\sqrt{r^2+d^2}}.
\end{split}
\end{align}
The required integration timescales to resolve uplift ($\tau_1$) and mass redistribution ($\tau_m$) as a function of the distance to the magma chamber and the volume change rate $\Delta V/\Delta \tau$ is given by Eq. \eqref{equ: integration time uplift} and \eqref{equ: integration time mass redistribution}, respectively.
For a current clock with $\sigma_0 = 3\times 10^{-16}$, the respective integration times are shown in Fig. \ref{fig:timescales_current}(a) and (b).
Resolving $\Delta U_\text{m}$ requires either long integration timescales of the order of a year or more or better clocks. However, for different geometries (e.g. flatter magma chambers) or more realistic models the mass redistribution term could be more significant.
For future clocks with $\sigma_0 = 1\times 10^{-17}$, the analogous plot is given in Fig. \ref{fig:timescales_future}.

\subsection{Ground displacement}
\label{groundUplift}
Treating the Earth as a perfect sphere, its gravitational potential is
\begin{align}
U(r, z) = - \frac{G M_\oplus}{\sqrt{r^2 + (R_\oplus-z)^2}},
\end{align}
where $M_\oplus$ is the mass and $R_\oplus$ is the radius of the Earth.

We expand the potential in a Taylor series around $(r,z)=(0,0)$ and
neglect all terms of order $\mathcal{O}(r^3),\mathcal{O}(z^3)$ and higher to obtain 
\begin{align}
U (r, z) &\approx - \frac{G M_\oplus}{R_\oplus} 
- \frac{G M_\oplus}{R_\oplus^2} z
+ \frac{G M_\oplus}{2 R_\oplus^3} (r^2 - 2 z^2)
\end{align}
The gravitational acceleration is given by
\begin{align}
g(r, z) = \frac{G M_\oplus}{r^2 + (R_\oplus-z)^2},
\end{align}
with the expansion
\begin{align}
g(r, z) &\approx \frac{G M_\oplus}{R_\oplus^2}
+ \frac{2 G M_\oplus}{R_\oplus^3} z
- \frac{G M_\oplus}{R_\oplus^4} (r^2 - 3 z^2) ~.
\label{equ:change grav acc Earth}
\end{align}

The linear term in the expansion is known as the free-air correction; we can neglect the higher orders. We define the gravitational acceleration at the surface as $g \equiv G M_\oplus / R_\oplus^2 \approx 9.81 \textrm{m/sec}^2$.
For a Mogi source $z = w$ (see Eq. \ref{equ:surface displacements}). The potential and gravity changes are
\begin{align}
\Delta U_1 &= - g \, z = g (1-\nu) \frac{\Delta V}{\pi} \frac{d}{(r^2+d^2)^{3/2}}, \\
\Delta g_1 &= 2 g \frac{z}{R_\oplus} = - 2 g (1-\nu)\frac{\Delta V}{\pi \, R_\oplus} \frac{d}{(r^2+d^2)^{3/2} }~.
\end{align}

\subsection{Direct signal}
\label{sub:Direct signal}

If there is inflow or outflow of magma, crustal rock of the volume $\Delta V$ will be replaced by magma, or vice versa.
This change in density gives a direct signal, due to the mass change $\Delta M = \Delta V (\rho_m - \rho)$, where $\rho_m$ and $\rho$ are the densities of magma and crustal rock, respectively. The resulting change in the gravitational potential from the presence of the density anomaly only is
\begin{align}
\Delta U_2 = - \frac{G \Delta M}{R}
 = - G (\rho_m - \rho)  \Delta V \frac{1}{\sqrt{d^2 + r^2}},
\end{align}
and the change in the gravitational acceleration is
\begin{align}
\Delta g_2 = \frac{\partial \Delta U_2}{\partial d}
= G (\rho_m - \rho)  \Delta V \frac{d}{(d^2 + r^2)^{3/2}}.
\end{align}


\subsection{The potential of the uplifted rock}
\label{sub:The potential of the uplifted rock}

Before the inflation of the cavity, the surface of the half-space is flat. But after the inflation there will be a hat of material peaking above the location of the cavity; this is where the clock and the gravimeter are located.
Obviously, this hat of material affects the outcome of a measurement.
As a simplification, most authors replace the hat by an infinite disc of height $d_0$. While this gives the correct first order result for the gravitational acceleration, the expression for the gravitational potential diverges.

In this section we discuss the gravitational potential and acceleration of this additional hat without making the disc approximation. This allows us to calculate the effect on clock measurements and to compute higher order corrections to the gravitational acceleration.

We denote the uplift directly above the cavity by
\begin{align}
d_0 \equiv |w(r=0)| = (1-\nu) \frac{\Delta V}{\pi} \frac{1}{d^2}  ~.
\end{align}
We shift the coordinates slightly to put the observer at $z = 0$. Thus, $z = d_0$ now corresponds to the surface of the ground before the uplift. The $z$ coordinate of the new surface at a given distance $r$ is
\begin{align}
\begin{split}
z(r) &= d_0 - |w(r)|\\
&= d_0 \left[1 - \left(\frac{r^2}{d^2}+1\right)^{-3/2} \right].
\end{split}
\end{align}
This can be inverted to express $r$ as a function of $z$:
\begin{align}
\label{equ:r for given surface displacement}
\left( \frac{r_z}{d} \right)^2
= \left( 1 - \frac{z}{d_0} \right)^{-2/3}-1.
\end{align}
The $z$ index is used to emphasize that $r_z$ is the distance at which the surface has altitude $z$.

In general, the gravitational acceleration and the gravitational potential of an observer located at $(r,z)=(0,0)$ due to the density distribution $\rho(r,z)$ are
\begin{align}
g_3 &= 2 \pi G\int_{-\infty}^\infty \int_0^\infty \frac{\rho(r,z) z}{(r^2+z^2)^{3/2}} r \, dr \, dz,
\\
U_3 &= - 2 \pi G \int_{-\infty}^\infty \int_0^\infty \frac{\rho(r,z)}{\sqrt{r^2+z^2}} r \, dr \, dz.
\end{align}
Assuming constant density, the acceleration $\Delta g_3$ and the potential $\Delta U_3$ due the uplifted mass are
\begin{align}
\Delta g_3 &= 2 \pi G \rho \int_0^{d_0} \int_0^{r_z} \frac{r z}{(r^2+z^2)^{3/2}} dr \, dz,
\\
\Delta U_3 &= - 2 \pi G \rho \int_0^{d_0} \int_0^{r_z} \frac{r}{\sqrt{r^2+z^2}} dr \, dz.
\end{align}
Performing the integrals over $r$ yields
\begin{align}
\Delta g_3 &= -2 \pi G \rho \int_0^{d_0} \left( \frac{z}{\sqrt{r_z^2 + z^2}} - 1 \right) dz,
\\
\Delta U_3 &= - 2 \pi G \rho \int_0^{d_0} \left( \sqrt{r_z^2 + z^2} - z \right) dz,
\end{align}
and defining a new variable $\zeta \equiv z/d_0$, we obtain
\begin{align}
\begin{split}
\Delta g_3
&= -2 \pi G \rho d_0 \int_0^{1} \left( \frac{\zeta}{\sqrt{\frac{r_z^2}{d_0^2} +\zeta^2}} - 1 \right) d\zeta \\
&= -2 \pi G \rho d_0 \int_0^{1}
\left(\frac{d_0}{d} \frac{\zeta}{\sqrt{ \left( 1 - \zeta \right)^{-2/3} + \frac{d_0^2}{d^2}\zeta^2 -1}} - 1 \right) d\zeta,
\end{split}
\\
\begin{split}
\Delta U_3
&= - 2 \pi G \rho \int_0^{d_0} \left( \sqrt{r_z^2 + z^2} - z \right) dz \\
&= - 2 \pi G \rho d_0 d \int_0^{d_0} \left( \sqrt{\left( 1 - \zeta \right)^{-2/3} + \frac{d_0^2}{d^2} \zeta^2 - 1} - \frac{d_0}{d} \zeta \right) d\zeta.
\end{split}
\end{align}
Because $d_0/d \ll 1$, we can expand around $d_0/d = 0$ and drop all terms that are of order $\mathcal{O}(d_0^2/d^2)$, then we can integrate and find
\begin{align}
\begin{split}
\Delta g_3 &\approx 2 \pi G \rho d_0 \left[ 1 - \left( 2-\frac{15\pi}{32} \right) \frac{d_0}{d} \right],
\\
\Delta U_3 &\approx - 2 \pi G \rho d_0 d \left[ 1 - \frac{1}{2} \frac{d_0}{d} \right].
\end{split}
\end{align}
To combine this with the other mass redistribution terms, we need the result as a function of $r$. In the infinite disk model in the literature, the gravitational acceleration at $r$ is taken to be that produced by a disk with a thickness equal to $|w(r)|$, the local value of the uplift,
\begin{align}
\begin{split}
\Delta g_3 &= 2 \pi G \rho |w(r)|
\\&= 2 G \rho (1-\nu) \Delta V \frac{d}{(r^2+d^2)^{3/2}} .
\end{split}
\end{align}
Integrating with respect to $d$ we obtain the corresponding potential
\begin{align}
\begin{split}
\Delta U_3
= - 2 G \rho (1-\nu) \Delta V \frac{1}{\sqrt{r^2+d^2}}.
\end{split}
\end{align}


\subsection{Gravitational potential due to density change}
\label{subsec:Gravitational potential due to density change}

In response to the inflating magma chamber, the density throughout the surrounding rock changes by $\Delta \rho(r,z)$, given by \eqref{equ: density change}.
The resulting change in the gravitational acceleration
\begin{align}
\Delta g_4(r) = - G (1-2\nu) \rho \Delta V \frac{d}{(r^2 + d^2)^{3/2}}. 
\end{align}
was calculated by \cite{Hagiwara1977}.
Following a similar approach, we determine the change in the gravitational potential due to this density variation.
This may be useful for the reader since Hagiwara's paper was written in Japanese.

First, a new function $K$ is defined
\begin{align}
K(x,y,z) = \left(x^2+y^2 + (z + d)^2\right)^{-3/2},
\end{align}
which allows to rewrite the density change \eqref{equ: density change} as
\begin{align}
\Delta\rho(x,y,z) = (1-2\nu) \rho \frac{\Delta V}{\pi} \left[ K(x,y,z) + (z+d) \frac{\partial K(x,y,z)}{\partial d} \right] ~.
\end{align}
The gravitational potential due to this change in density is
\begin{align}
\begin{split}
\Delta U_4(x,y,z) &= - \int_{\mathbb{R}^3} \frac{G \Delta \rho(x',y',z') dx' \,dy' \, dz'}{\sqrt{(x-x')^2+(y-y')^2+(z-z')^2}}
\\
&=\kappa \int_{\mathbb{R}} \left[1+(z'+d) \frac{\partial}{\partial d} \right] \Phi_K(x,y,z,z') dz',
\end{split}
\end{align}
where we defined
\begin{align}
\Phi_K(x,y,z,z') &\equiv \int_{\mathbb{R}^2} \frac{K(x',y',z') dx' \,dy'}{\sqrt{(x-x')^2+(y-y')^2+(z-z')^2}},
\\
\kappa &\equiv - G (1-2\nu) \rho \frac{\Delta V}{\pi}.
\end{align}
Further, we assume that before the inflation of the magma chamber the density $\rho$ is constant.

The 2D Fourier transform, assuming $z=0$, is
\begin{align}
\begin{split}
\Phi^*_{K}(k_x,k_y) &= \frac{1}{2\pi} \int_{\mathbb{R}^2} \Phi_{K}(x,y,0,z') e^{-i (k_x x+k_y y)} dx\,dy
\\
&= \left[ \int_{\mathbb{R}^2} K(x',y',z') e^{-i (k_x x'+k_y y')} dx' \,dy' \right]
\\&\qquad\times \left[ \frac{1}{2\pi} \int_{\mathbb{R}^2} \frac{e^{-i (k_x u+k_y v)}}{\sqrt{u^2+v^2+z'^2}} du \, dv \right]
\\
&= 2 \pi \frac{ e^{ -|z'+d|\sqrt{k_x^2 + k_y^2} } }{ z'+d }
\frac{ e^{ -|z'|\sqrt{k_x^2 + k_y^2} } }{ \sqrt{k_x^2 + k_y^2} } 
\\
&= 2 \pi \frac{ e^{ -(2 z' + d)\sqrt{k_x^2 + k_y^2} } }{ (z'+d) \sqrt{k_x^2 + k_y^2} }.
\end{split}
\end{align}
We substituted $u = x-x'$ and $v = y-y'$ and used $dx\,dy\, dx'\, dy' = du\, dv \,dx' \,dy'$. Further, we used the Fourier integrals
\begin{align}
\frac{1}{2\pi} \int_{\mathbb{R}^2} \frac{e^{-i(k_x x + k_y y)}}{(x^2+y^2+z^2)^{1/2}} dx\,dy
&= \frac{ e^{ -|z|\sqrt{k_x^2 + k_y^2} } }{ \sqrt{k_x^2 + k_y^2} },
\\
\frac{1}{2\pi} \int_{\mathbb{R}^2} \frac{e^{-i(k_x x + k_y y)}}{(x^2+y^2+z^2)^{3/2}} dx\,dy
&= \frac{ e^{ -|z|\sqrt{k_x^2 + k_y^2} } }{ z },
\end{align}
and we used that the density vanishes for $z'<0$.
Taking the Fourier transform of $\Delta U_4$
\begin{align}
\begin{split}
\Delta U_4^*(k_x,k_y)
&= \frac{1}{2\pi} \int_{\mathbb{R}^2} \Delta U_4(x,y) e^{-i (k_x x+k_y y)} dx\,dy
\\
&= \kappa \int_{\mathbb{R}} \left[1+(z'+d) \frac{\partial}{\partial d} \right]
\\&\quad\times \left( \frac{1}{2\pi} \int_{\mathbb{R}^2}  \Phi_{K}(x,y) e^{-i (k_x x+k_y y)} dx\,dy\right) dz'
\\
&= \kappa \int_{\mathbb{R}} \left[1+(z'+d) \frac{\partial}{\partial d} \right] \Phi^*_{K}(k_x,k_y) dz',
\end{split}
\end{align}
and using the expression for $\Phi^*_{K}$ above, we obtain
\begin{align}
\begin{split}
\Delta U_4^*(k_x,k_y)
&= 2 \pi \kappa \int_{\mathbb{R}} \left[1 + (z'+d) \frac{\partial}{\partial d} \right] \frac{ e^{ -(2 z' + d)\sqrt{k_x^2 + k_y^2} } }{ (z'+d) \sqrt{k_x^2 + k_y^2} } dz'
\\
&= -2 \pi \kappa \int_0^\infty e^{-(2z'+d)\sqrt{k_x^2 + k_y^2}} dz'
\\
&= - \pi \kappa \frac{e^{-d\sqrt{k_x^2+k_y^2}}}{\sqrt{k_x^2+k_y^2}},
\end{split}
\end{align}
where we used that outside the half-space (i.e. for $z'<0$) the density (contained in $\kappa$) vanishes.
The inverse Fourier transform is
\begin{align}
\begin{split}
\Delta U_4(x,y) &= \frac{1}{2\pi} \int_{\mathbb{R}^2} \Delta U_4^*(k_x,k_y) e^{i (k_x x+k_y y)} dk_x dk_y
\\
&= - \frac{1}{2} \kappa \int_{\mathbb{R}^2} \frac{e^{-d\sqrt{k_x^2+k_y^2}}}{\sqrt{k_x^2+k_y^2}} e^{i (k_x x+ky_ y)} dk_x dk_y,
\end{split}
\end{align}
and introducing polar coordinates $k_x = k \cos\theta, k_y = k \sin\theta$ (such that $dk_x dk_y = k \,dk\,d\theta$) and orientating the coordinate axes such that $(x,y) = (r,0)$, the integral can be written as
\begin{align}
\begin{split}
\Delta U_4(x,y)
&= - \frac{1}{2} \kappa \int_0^\infty \int_0^{2\pi} \frac{e^{-d k}}{k} e^{i k r \cos\theta} k \,d\theta \, dk
\\
&= - \frac{1}{2} \kappa \int_0^\infty e^{-d k} \left( \int_0^{2\pi} e^{i k r \cos\theta} \,d\theta \right) dk
\\
&= - \kappa \frac{\pi}{\sqrt{r^2 + d^2}}.
\end{split}
\end{align}
Here, we used the definition of the Bessel functions and the integral 6.751(3) in \cite{integral.table}.
Finally, we find
\begin{align}
\Delta U_4(r) = G (1-2\nu) \rho \Delta V \frac{1}{\sqrt{r^2 + d^2}}.  
\end{align}


\subsection{Optimizing the placement of a second clock}
\label{optimalClock}

We now consider the question of the optimal location for measuring the depth and volume change of a magma chamber. We will assume that one clock is directly over the magma chamber. What is the optimal location of a second clock? Recall that there must also be a distant reference clock. For this example, we assume that the horizontal location of the magma chamber is known from a previous survey. If this were not known, it could also be measured, but more clock locations would be required.

The shift in the gravitational potential at clock $i$ can be written
\begin{align}
\Delta U_{i}(r) &= C_0 \Delta V \frac{d}{(r_{i}^2+d^2)^{3/2}} ~, \\
C_0 &\equiv \frac{G M_\oplus}{R_\oplus^2}\frac{1-\nu}{\pi},
\end{align}
where $d$ is the depth of the magma chamber, $r_{i}$ the horizontal distance of clock $i$ from the magma chamber. We assume that $\nu$ is known, thus $C_0$ is a constant.

We measure two quantities, the differences between each of the clocks and the reference clock. These are
\begin{align}
\Delta U_1 &= C_0 \Delta V \frac{1}{d^2}, \\
\Delta U_2 &= C_0 \Delta V \frac{d}{(r^2 + d^2)^{3/2}} ~.
\end{align}

The measurements of each of these quantities is affected by clock noise. The three clocks will have uncorrelated noise, but the $\Delta U_{i}$ will be correlated because they each depend on the reference clock. We write the variance of the two clocks near the magma chamber as $\sigma^2_{M}$, and the variance of the reference clock as $\sigma^2_{R}$. Then the variances and covariance of $\Delta  U_1$ and $\Delta U_2$ are
\begin{align}
\mathrm{Var}(\Delta U_1) = \sigma^2_{M} + \sigma^2_{R}, \\
\mathrm{Var}(\Delta U_2) = \sigma^2_{M} + \sigma^2_{R}, \\
\mathrm{Cov}(\Delta U_1, \Delta U_2) = \sigma^2_{R} ~.
\end{align}

The noise variance in the measurement of $d$ is
\begin{align}
\begin{split}
\mathrm{Var}(d) = \left(\frac{\partial d}{\partial \Delta U_1}\right)^2 \mathrm{Var}(\Delta U_1)
+ \left(\frac{\partial d}{\partial \Delta U_2}\right)^2 \mathrm{Var}(\Delta U_2) \\
+ 2 \frac{\partial d}{\partial \Delta U_1} \frac{\partial d}{\partial \Delta U_2} \mathrm{Cov}(\Delta U_1, \Delta U_2)
\end{split}
\end{align}
and the same holds replacing $d$ with $\Delta V$.
It is easiest to calculate the partial derivatives by calculating first the partial derivatives of the $\Delta U_i$ with respect to $d$ and $\alpha$, and inverting. They are
\begin{align}
\frac{\partial \Delta V}{\partial \Delta U_1} &= \frac{d^2 (r^2 - 2 d^2)}{3 C_0 \, r^2}, \\
\frac{\partial \Delta V}{\partial \Delta U_2} &= \frac{2 (r^2 + d^2)^{5/2}}{3 C_0 d \, r^2}, \\
\frac{\partial d}{\partial \Delta U_1} &= - \frac{d^3 (r^2 + d^2)}{3 C_0 \Delta V \, r^2}, \\
\frac{\partial d}{\partial \Delta U_2} &= \frac{(r^2 + d^2)^{5/2}}{3 C_0 \Delta V \, r^2} ~.
\end{align}

We find the optimal location for the second clock by minimizing the variance of either $\Delta V$ or $d$, as a function of $r$, the horizontal distance to the second clock. Fortunately, we find that $d$ and $\Delta V$ are optimized for similar values of $r$, and the minima are fairly broad. A reasonable compromise, in the case where the reference clock has the same performance as the other clocks, is $r \approx 0.78 \, d$. With this clock placement, we have 
\begin{align} 
\mathrm{Var}(\Delta V) \approx \left( 4.6 \, \frac{d^2}{C_0} \right)^2 \sigma_{M}^2, \\
\mathrm{Var}(d) \approx \left( 2.2 \, \frac{d^3}{C_0 \Delta V} \right)^2 \sigma_{M}^2.
\end{align}
We can summarize this more succinctly in terms of fractional errors. We define $\delta h \equiv \sigma_M / g$, which is the standard deviation of the measurement of a single clock, in terms of equivalent height. The maximum value of the uplift (at the summit) is $h = C_0 \Delta V / (g d^2)$. The standard deviation of one clock is then $\sigma_{M} = (C_0 \Delta V / d^2) \delta h / h$. Given this, the fractional accuracies of our measurements are
\begin{align}
\frac{\delta \Delta V}{\Delta V} &\approx 4.6 \, \frac{\delta h}{h}, \\
\frac{\delta d}{d} &\approx 2.2 \, \frac{\delta h}{h}.
\end{align}
A frequency stability of $10^{-18}$ corresponds to $1$ cm, which is achieved with current clocks after an integration of 7 hours. If each clock has this performance and the maximum uplift is $10$ cm, this is a fractional accuracy of $10 \%$, giving a fractional accuracy on $d$ of $22 \%$ and on $\Delta V$ of $46 \%$. If instead, we integrate for a month, the fractional accuracies improve by an order of magnitude leading to fractional accuracies of $2.2\%$ and $4.6\%$, respectively.

In the case of a reference clock that is much better than the others ($\sigma_R \ll \sigma_M$), the numbers become $r \approx 0.87 d$, $\delta \Delta V / \Delta V \approx 3.7 \, \delta h / h$, and $\delta d / d \approx 2.0 \, \delta h / h$.

\section{Solid Earth tides}
\label{subsec:Solid Earth tides}
Throughout this section we follow \cite{agnew} and we use standard spherical coordinates.
The gravitational potential of an external body can be written as a multipole expansion
\begin{align}
\begin{split}
U_\text{ext} 
&= -\frac{G M_\text{ext}}{\rho}
\\
&= -\frac{G M_\text{ext}}{R} \sum_{n=0}^\infty \left(\frac{r}{R}\right)^n P_n(\cos\alpha).
\end{split}
\end{align}
Here, $\rho$ is the distance between the external body and the location of the clock, $R$ is the distance between the body and the Earth's centre of mass, $r$ is the distance between the Earth's centre and the clock and $\alpha$ is the angle between the two vectors pointing from the centre of the Earth to the external mass and the clock. The $P_n$ denote the Legendre polynomials.
In the sum, the $n=0$ term can be neglected since it is constant and therefore does not contribute to the force since it will drop out when taking the gradient. The $n=1$ term $(U_\text{ext,n=1} = -G M_\text{ext}/R^2 r\cos\alpha)$  causes the orbital acceleration and therefore, by the definition of tides, the tidal potential is
\begin{align}
U_\text{tid}(r,\alpha)
= -\frac{G M_\text{ext}}{R} \sum_{n=2}^\infty \left(\frac{r}{R}\right)^n P_n(\cos\alpha).
\end{align}
By $U_\text{tid}^n$ we will denote the n-th order component of the expansion.

The tidal deformation is mainly caused by the Moon and the Sun; the other planets cause very minor effects.
Since typically $r/R$ is small (for the Moon: $r/R \approx 1/60$, Sun: $r/R \approx 1/23000$), it is sufficient to just consider the first few orders of the expansion.

If the Earth was a completely rigid spherical body with no external mass acting tidally, the gravitational acceleration at the surface would be a constant $g \equiv G M_\oplus/R_\oplus^2$. Changing the radial distance by a small amount $\Delta r$ would change $g$ by 
$\Delta g = - 2 G M_\oplus/R_\oplus^3 \Delta r$.
Thus, going farther away ($\Delta r > 0$) weakens the gravitational acceleration.

First, we consider just the effect of an external body, letting the Earth be perfectly rigid. This corresponds to the Love numbers (introduced below) being zero.
The value of the tidal potential on the surface is $U_\text{tid}(R_\oplus,\alpha)$.
This induces a change in the gravitational acceleration at the surface of
\begin{align}
\label{equ:tidal grav acceleration}
\begin{split}
\Delta g(\alpha) &= - \left. \frac{\partial U_\text{tid}(r,\alpha)}{\partial r} \right|_{r=R_\oplus}
\\
&= - \sum_{n=2}^\infty n \frac{U_\text{tid}^{(n)}(R_\oplus,\alpha)}{R_\oplus}.
\end{split}
\end{align}
Just taking the first ($n=2$) term, this is
\begin{align}
\begin{split}
\Delta g(\alpha) &= - 2 \frac{U_\text{tid}^{(2)}(R_\oplus,\alpha)}{R_\oplus}
\\
&= \frac{G M_\text{ext}}{R^3} R_\oplus ( 3 \cos^2\alpha - 1).
\end{split}
\end{align}
Thus, there is an outward pull, $\Delta g>0$, at the sides of the Earth facing 
($\alpha = 0$) and opposing ($\alpha = \pi$) the external body. Halfway between, there is an inward pull ($\alpha = \pi/2, 3\pi/2$).
Since the Earth is not a rigid body, it is deformed by tidal forces. Its tidal response is quantified by the Love numbers $k_n$ and $h_n$; there is also a Love number $l_n$ which we do not use here. The larger the Love numbers, the stronger the deformation.

The gravitational acceleration arising from the tidal interaction has three components.
First, there is the direct effect from the external body \eqref{equ:tidal grav acceleration}.
Second, the deformation of the Earth gives rise of an additional gravitational potential $k_n U^{(n)}_\text{tid}$, resulting in an additional acceleration.
Third, due to the deformation, the altitude of the observer changes by $- h_n U^{(n)}_\text{tid}/g$ compared to a non-deformed Earth.
Thus, the effective tidal potential measured by a clock on the surface, for a given $n$, is
\begin{align}
U^{(n)}_\text{tid,eff} = (1 + k_n - h_n) U^{(n)}_\text{tid} \label{equ:eff tidal pot} ~.
\end{align}

The potential of the deformed Earth is $k_n U_{(n)}$. It can be written as an expansion around its value $U_\text{tid}(R_\oplus,\alpha)$ at the undistorted surface.
Since the gravitational field satisfies Poisson's equation, it is a harmonic function outside the Earth and therefore, we can write it as an expansion of the form
\begin{align}
U_\text{Earth def}(r,\alpha)
= \sum_{n=0}^\infty \left( \frac{R_\oplus}{r} \right)^{n+1} F_n(\alpha),
\end{align}
where the functions $F_n(\alpha)$ have to be chosen such that they match $U_\text{tid}(R_\oplus,\alpha)$ at the Earth's surface, giving
\begin{align}
U_\text{Earth def}(r,\alpha)
= - \frac{G M_\text{ext}}{R} \sum_{n=2}^\infty \left(\frac{R_\oplus}{R}\right)^n \left( \frac{R_\oplus}{r} \right)^{n+1} P_n(\cos\alpha).
\end{align}
The tidal acceleration is now given by
\begin{align}
\begin{split}
&- \left. \frac{\partial ( U_\text{tid}+k_n U_\text{Earth def} )}{\partial r} \right|_{r=R_\oplus}
\\
&= \frac{G M_\text{ext}}{R} \sum_{n=2}^\infty \frac{1}{R_\oplus} \left(n - k_n (n+1) \right) \left(\frac{R_\oplus}{R}\right)^n
\\&\qquad \times P_n(\cos\alpha),
\end{split}
\end{align}
which can be written as
\begin{align}
\begin{split}
- \sum_{n=2}^\infty n \left(1 - \frac{n+1}{n} k_n \right) \frac{U_\text{tid}^n(R_\oplus,\alpha)}{R_\oplus}.
\end{split}
\end{align}
The change in the gravitational acceleration due to a small radial displacement $\Delta r$ close to the surface of the Earth is
\begin{align}
- \frac{2 G M_\oplus}{R_\oplus^3} \Delta r
= - 2 \frac{G M_\oplus}{R_\oplus^2} \frac{\Delta r}{R_\oplus}
= - 2 g \frac{\Delta r}{R_\oplus}.
\end{align}
The final effective tidal acceleration on the surface is
\begin{align}
\Delta g^{(n)}_\text{tid,eff} = \left(n - (n+1) k_n + 2 h_n \right) \frac{U_\text{tid}^n(R_\oplus,\alpha)}{R_\oplus} \label{equ:eff tidal acc} ~.
\end{align}

From \eqref{equ:eff tidal pot} and \eqref{equ:eff tidal acc}, we see a key feature of clocks relative to gravimeters. Gravimeters are relatively more sensitive to effects at higher $n$, i.e. shorter wavelengths, than are clocks. This is because they measure a spatial derivative of the potential rather than the potential itself. In contrast, the response to uplift only depends on motion in the entire gravitational field of the Earth, and so the coefficients of $h_n$ do not depend on $n$.

Note that clocks are best at constraining the lower order multipoles. Instruments that measure higher order derivatives of the geopotential (e.g. gradiometers) are more sensitive to the higher order multipoles. Clocks will only be useful for measuring the $n=2$ and possibly $n=3$ tides. However, a detailed analysis of the multipole structure based on clock data is beyond the subject of this paper.

\label{lastpage}

\end{document}